\documentclass[preprint]{revtex4}

\input epsf
\usepackage{amsmath,amssymb}
\usepackage{graphicx}

\begin{document}
\titlepage
\title{Codimension Two Branes in Einstein-Gauss-Bonnet Gravity}
\author{Peng Wang$^1$\footnote{E-mail: pewang@eyou.com} and Xin-He Meng$^{1,2}$
\footnote{E-mail: xhmeng@phys.nankai.edu.cn}} \affiliation{1. Department of Physics,
Nankai University, Tianjin, 300071, P.R.China \\ 2. CCAST (World
Laboratory) P.O. Box 8730, Beijing 100080, China}

\begin{abstract}
Codimension two branes play an interesting role in attacking the
cosmological constant problem. Recently, in order to handle some
problems in codimension two branes in Einstein gravity, Bostock
{\it et al.} have proposed using six-dimensional
Einstein-Gauss-Bonnet (EGB) gravity instead of six-dimensional
Einstein gravity. In this paper, we present the solutions of
codimension two branes in six-dimensional EGB gravity. We show
that Einstein's equations take a "factorizable" form for a
factorized metric tensor ansatz even in the presence of the
higher-derivative Gauss-Bonnet term. Especially, a new feature of
the solution is that the deficit angle depends on the brane
geometry. We discuss the implication of the solution to the
cosmological constant problem. We also comment on a possible
problem of inflation model building on codimension two branes.
\end{abstract}

\maketitle

\section{Introduction}

The idea of braneworlds and large extra dimensions \cite{add}
implies that the cosmological constant problem (see
Refs.\cite{carroll-cc, weinberg} for reviews) may be a clue of our
unawareness of the true nature of spacetime: vacuum energy may be
large, but it simply does not gravitate in the 4-dimensional
braneworld where we are living in. The key point is that,
cosmological constant is a reflection of 4-dimensional spacetime
geometry and thus is what is observed directly in cosmological
observations. The puzzle arises only after we use General
Relativity to find that cosmological constant describes vacuum
energy of standard model particles. So if we modify gravity theory
by introducing higher dimensional spacetime and objects like
branes, it is possible that 4-dimensional cosmological constant is
not linked to 4-dimensional vacuum energy, but something else such
as higher dimensional vacuum energy. See, e.g., Refs.\cite{dval,
self} for some earlier endeavors in this direction.

Recently, Carroll and Guica presented an interesting exact
solution of this type \cite{carroll-foot}. They considered a
factorizable braneworld spacetime with two extra dimensions and
explicit brane sources. The compactification manifold has the
topology of a two-sphere, and is stabilized by both a bulk
cosmological constant and a magnetic flux. From their solution,
they found that the flat nature of the 4-dimensional geometry is
independent of the brane tension. This feature moves the
cosmological constant problem completely into the extra
dimensions. Of course, this is not a complete solution to the
cosmological constant problem, since it still needs fine-tuning in
the bulk. But it transforms the nature of the problem in a
suggestive way.

The interesting feature of Carroll and Guica's solution is not an
accident (see Ref.\cite{codim2} for some other models of
codimension 2 branes that share the similar feature, see also
Ref.\cite{gibbons} for earlier ideas along this line). It can be
shown that the independence of 4-dimensional geometry on the brane
tension is a general feature of codimension 2 branes in
factorizable spacetime in Einstein gravity. The following
discussions will also be helpful for us to understand the
properties of codimension 2 branes in Einstein-Gauss-Bonnet
gravity (see Sec.\ref{egb}).

Let's consider a factorizable metric ansatz,
\begin{equation}
ds^2=G_{AB}dX^A dX^B=g_{\mu\nu}(x)dx^\mu dx^\nu+\gamma_{ab}(y)dy^a
dy^b\label{metric}
\end{equation}
where $A,B=0,...,5$, $\mu,\nu=0,...,3$ and $a,b=4,5$.  The
$\gamma_{ab}$ is the metric of a Einstein manifold with curvature
$k=-1, 0, 1$. Note that due to the presence of branes, there will
be deficit angles in the extra dimensions at the positions of the
branes (see Sec.\ref{solution}), but this will not influence the
local geometry of the extra dimensions at other points.

We will consider the simplest model of branes which is also the
case considered in most of the literature on braneworld cosmology:
the branes are described by Nambu-Goto action (see
Ref.\cite{carter} for an elegant review),
\begin{equation}
S_{NG}=\int d^6X\sqrt{|G|}\mathcal{L}_{brane}\ ,\label{NG}
\end{equation}
where
\begin{equation}
{\mathcal{L}}_{brane}=-\sum_i\int d^4x\sqrt{{|g|\over
|G|}}\sigma_i\delta^{(6)}(X-X_i(x))\ ,\label{}
\end{equation}
in which $i$ labels the branes, $\sigma_i$ and $X_i$ are the
tension and position of the $i$th brane, respectively. The
energy-momentum tensor of branes follow by varying $G^{AB}$ in
(\ref{NG}) \cite{carter}
\begin{equation}
  T_{AB}^b = -\sum_i{\sigma_i\over\sqrt{\gamma}}
  \left(\begin{matrix} g_{\mu\nu}  & 0 \cr 0 & 0 \end{matrix}\right)
  \delta^{(2)}(y-y_i)\ ,
\end{equation}

With the help of the fact that the Einstein tensor,
$G_{ab}=R_{ab}-\frac{1}{2}h_{ab} R$, vanishes identically for any
2-dimensional metric $h_{ab}$ and denoting the bulk
energy-momentum tensor by $T^B_{AB}$, contracting the transverse
component of the Einstein equation gives
\begin{equation}
R[g]=-\frac{2}{M_6^4}T^B_2\ ,\label{1.1}
\end{equation}
where $T^B_\gamma\equiv T^B_{ab}\gamma^{ab}$; while contracting
the longitudinal component gives
\begin{equation}
R[\gamma]+{1\over2}R[g]=-{1\over2M_6^4}[T^B_g+T^b]\ .\label{1.2}
\end{equation}
where $T^B_g\equiv T^B_{\mu\nu}g^{\mu\nu}$ and $T^b\equiv
T^b_{\mu\nu}g^{\mu\nu}$. Now, Eq.(\ref{1.1}) tells us that
\emph{the scalar curvature of the 4-dimensional spacetime is
totally determined by the transverse component of the total bulk
energy-momentum tensor}. Thus if we assume the 4-dimensional
geometry to be maximal symmetric, then it is determined totally by
the transverse part of the bulk energy momentum tensor.
Specifically, the 4-dimensional geometry does not depend on the
brane tension. Then, after we find $R[g]$ from Eq.(\ref{1.1}),
substituting it into Eq.(\ref{1.2}), we can find the bulk
curvature $R[\gamma]$. In sum, for codimension 2 branes in
factorizable spacetime, the brane geometry is determined by the
\emph{transverse} component of the Einstein equations and the bulk
geometry is determined by the \emph{longitudinal} component of the
Einstein equations. Roughly speaking, we can say that Einstein
equations in factorizable spacetime are also ``factorizable". This
is the secret of codimension 2 branes in Einstein gravity.

While the above discussion is exiting, unfortunately, when
considering realistic cosmological evolution of this model, we
will encounter some fundamental difficulties. One of them is that
if we assume the brane energy-momentum tensor to be of the form as
the perfect fluid, i.e. $T^\mu_\nu=\{\rho, p,p,p\}\delta^\mu_\nu$,
then $\rho$ and $p$ must satisfy $\rho+p=0$, i.e. it behaves like
the brane tension \cite{cline}. This forbids us adding dust and
radiation on the brane, thus it is cosmologically unrealistic. To
remedy this and other difficulties of codimension 2 branes in
Einstein gravity, recently, Bostock \emph{et al.} suggested that
we may add the Gauss-Bonnet term to the 6-dimensional
gravitational action \cite{bgns} (however, see also
Ref.\cite{cline2} and reference therein for some other suggestions
to handle this problem). It is also worth commenting that the idea
that 6-dimensional Einstein-Gauss-Bonnet (EGB) gravity could be
relevant in relation to the cosmological constant problem was
originally presented in Ref.\cite{Iglesias} (see also
Ref.\cite{series} for some subsequent related works). The
Gauss-Bonnet term is quadratic in the curvature tensors and is a
topological invariant in 4-dimensional manifold (see, e.g.,
Ref.\cite{dm}); but in higher dimensions, it has the well-know
property that the equation of motion derived from it remains
second order differential equations of the metric. Furthermore,
considering higher derivative terms is also necessary to develop
the braneworld scenario in a more string theoretic setting (see,
e.g., Ref.\cite{Odintsov-hd}). Specifically, the Gauss-Bonnet
combination arises as the leading order for quantum corrections in
the heterotic string effective action and is the only quadratic
combination of curvature tensors that is ghost-free \cite{GB}.

Thus, the investigation of codimension 2 branes in EGB gravity is
well motivated (see Ref.\cite{recent} for some other recent
discussion of codimension 2 branes in EGB gravity). Of course, one
of the best way to understand the property of a gravity theory is
studying its exact solutions. Especially in the present case, the
EGB gravity is intended to remedy the model in Einstein gravity.
Thus one natural step is to derive and compare the corresponding
solutions in EGB gravity under the same assumption of spacetime
geometry and matter content with Einstein gravity case. In
particular, it is important to check that the important property
in Einstein gravity, i.e. the independence of the 4-dimensional
geometry on the brane tension, is retained in EGB gravity. If this
were not the case, considering EGB gravity would not be so well
motivated. We will see in Sec.\ref{egb} that the discussion above
for Einstein gravity also applies to EGB gravity, thus EGB gravity
retains the main features of Einstein gravity. In
Sec.\ref{solution}, we will also see that some new features will
arise in EGB gravity. The last section, Sec.\ref{CD}, is devoted
to conclusions and we comment on inflation model building in the
codimension 2 brane scenario.

\section{Einstein-Gauss-Bonnet equation in factorizable spacetime}
\label{egb}

Let's consider adding the Gauss-Bonnet term to modify the
6-dimensional gravity \cite{bgns}, which is described by the
action
\begin{equation}
S_6=\int d^6X \sqrt{|G|}{M_6^4\over 2}[R+\alpha R_{GB}^2]\
,\label{gbaction}
\end{equation}
where $\alpha$ is the Gauss-Bonnet coupling constant with
dimension $[\alpha]=(mass)^{-2}$. Following the original
derivation \cite{GB}, one generally assumes $\alpha\ge0$, but in
the literature the $\alpha<0$ case is also often discussed. We
will see in Sec.\ref{solution} that, from the exact solution we
found, the requirement of the geometry to be nonsingular will rule
out a negative Gauss-Bonnet coupling constant. The Gauss-Bonnet
term $R_{GB}$ is given by
\begin{equation}
R^2_{\rm GB}=R^2
    -4R^{AB}R_{AB}+ R^{ABCD}R_{ABCD} \ .
    \label{GBterm}
\end{equation}

Then the gravity field equation in 6-dimensional is described by
the Einstein-Gauss-Bonnet equation,
\begin{equation}
G_{AB}+\alpha H_{AB}={1\over M_6^4}T_{AB} \ ,\label{EGB}
\end{equation}
where
\begin{equation}
H_{AB}=-\frac{1}{2}g_{AB} R_{GB}^2 + 2 R R_{AB}
    - 4 R_{AC} R^C{}_B  -  4 R^{CD} R_{ACBD} + 2 R_A{}^{CDE} R_{BCDE}\ .  \
\end{equation}

While the EGB equation (\ref{EGB}) is rather complicated, it can
be shown that in factorizable spacetime, the EGB equation can be
simplified into a rather illuminating form: after inserting the
ansatz (\ref{metric}) into the EGB equations (\ref{EGB}), the
transverse and longitudinal EGB equations can be simplified to
give
\begin{equation}
\alpha R_{GB}^2[g]+R[g]=-{1\over M_6^4}T^B_\gamma\
,\label{transverse}
\end{equation}
\begin{equation}
(\alpha R[g]+1)R[\gamma]+{1\over2}R[g]=-{1\over
2M_6^4}[T^B_\gamma+T^b]\ .\label{longitude}
\end{equation}
From those two equations we can see that the main feature of
codimension 2 branes in Einstein gravity is retained in EGB
gravity: the scalar curvature of the 4-dimensional spacetime is
still determined only by the transverse component of the bulk
energy-momentum tensor from the transverse EGB equation
(\ref{transverse}); the bulk geometry is then determined by the
longitudinal equation (\ref{longitude}). So the EGB equations are
still ``factorizable" in factorizable spacetime. Thus in the EGB
gravity, we still can move the cosmological constant problem
completely into the bulk.

Now we have good motivation to proceed to see how the spacetime
solutions will be modified in EGB gravity. As a first remark, it
is interesting to see from Eqs.(\ref{transverse}) and
(\ref{longitude}) that the Gauss-Bonnet term couples only with the
4-dimensional scalar curvature $R[g]$. Thus, when the
4-dimensional geometry is flat, EGB equations will always reduce
to Einstein equations (\ref{1.1}) and (\ref{1.2}). So in this
case, the bulk solution is the same as the one given in
Ref.\cite{carroll-foot}. Thus what is really interesting is the
case when the brane geometry is not flat. In the next section, we
will consider de Sitter geometry on the brane.

\section{De Sitter branes in Einstein-Gauss-Bonnet gravity}
\label{solution}

From the recent cosmological observation that our universe is
currently accelerating \cite{obs}, we are interested in solutions
with de Sitter geometry on the brane. So we will consider in this
section the case that the geometry on the brane is de Sitter, i.e.
$R[g]_{\mu\nu}=\Lambda_4g_{\mu\nu}$ and $R[g]=4\Lambda_4$, where
$\Lambda_4\ge0$ is the 4-dimensional cosmological constant. Under
those assumptions of spacetime geometry, it can be seen from
Eqs.(\ref{transverse}) and (\ref{longitude}) that the bulk
energy-momentum tensor $T^B_{AB}$ must be constant along the bulk.

Let's first discuss the 4-dimensional geometry by
Eq.(\ref{transverse}). Under the assumption of maximal symmetric,
it can be rewritten as
\begin{equation}
{8\over3}\alpha\Lambda_4^2+4\Lambda_4=-{1\over M_6^4}T^B_\gamma\
,\label{3.1}
\end{equation}

From Eq.(\ref{3.1}), we can find that the 4-dimensional
cosmological constant is given in terms of the bulk
energy-momentum tensor by
\begin{equation}
\Lambda_4={3\over4\alpha}\left
[-1\pm\sqrt{1-{2\alpha\over3M_6^4}T^B_\gamma}\ \right]\ .
\label{3.3}
\end{equation}
Thus, the first different feature we encounter in the EGB gravity
is that, for any given $T^B_{AB}$, unless it satisfies
$T^B_\gamma={3M_6^4\over2\alpha}$, we will have two solutions of
the brane geometry. After the brane geometry is determined, the
bulk geometry is uniquely determined by the brane geometry from
Eq.(\ref{longitude}). Thus, generally, for any given bulk matter
content, there will be two different solutions of the EGB
equation. This is obviously not a pleasant feature. However, we
will argue that the ``-" branch of the solution is unphysical and
should be discarded. It can be seen from Eq.(\ref{3.3}) that for
the ``-" branch, the coefficient of $R[\gamma]$ in
Eq.(\ref{longitude}), i.e. $4\alpha\Lambda_4+1$, is always
negative; while in Einstein gravity, i.e. $\alpha=0$, it is always
positive. This means that for the ``-" branch, the gravity in the
transverse dimension is \emph{repulsive}: positive bulk energy
density will give rise to negative curvature and only negative
tension branes can give rise to a positive deficit angle. We think
those properties are too exotic so should be regarded as
unphysical. Thus in the following discussions, we will discard the
``-" branch.

Then let's discuss the bulk geometry from Eq.(\ref{longitude}),
which now can be written as
\begin{equation}
(4\alpha\Lambda_4+1)M_6^4R[\gamma]=-{1\over2}T^B_g
-2M_6^4\Lambda_4+{2\sigma\over\sqrt{|\gamma|} }\delta^{(2)}(y)\
.\label{3.2}
\end{equation}

First, in the case of a vacuum bulk, i.e. $T^B_{AB}=0$. From the
``+" branch of Eq.(\ref{3.3}), we have $\Lambda_4=0$, and
Eq.(\ref{3.2}) will just reduce to Einstein gravity. Thus the bulk
geometry will be same as the case discussed in Ref.\cite{sundrum}.

Next, let's consider the presence of bulk fields. Following
Refs.\cite{carroll-foot, codim2, sta}, we expect the extra
dimensions to have the topology of a sphere $\mathcal{S}^2$. Thus
the 2-dimensional metric $\gamma_{ab}$ will be of the form,
\begin{equation}
\gamma_{ab}dy^ady^b=a_0^2(d\theta^2+\beta^2\sin^2\theta
d\varphi^2)\ ,\label{2dmetric}
\end{equation}
where $a_0$ is the size of the extra dimensions and $\beta$ is
related to the deficit angle $\delta$ by $\delta=2\pi(1-\beta)$.

Transforming the metric (\ref{2dmetric}) into the conformal form,
\begin{equation}
\gamma_{ab}dy^ady^b=\psi(r)(dr^2+r^2d\varphi^2)\ ,\label{}
\end{equation}
where $\psi$ is given by \cite{carroll-foot}
\begin{equation}
\psi(r) = {4\beta^2 a_0^2\over
  r^2[(r/r_0)^\beta + (r/r_0)^{-\beta}]^2}\ ,\label{1.6}
\end{equation}
and substituting this into Eq.(\ref{3.2}), it can be found that
$a_0$ and $\beta$ are given by
\begin{equation}
a_0^2=\frac{M_6^4(1+4\alpha\Lambda_4)}{-{1\over4}T^B_g-M_6^4\Lambda_4}\
, \label{bulk1}
\end{equation}
\begin{equation}
\beta=1-\frac{\sigma}{2\pi M_6^4(1+4\alpha\Lambda_4)}\
.\label{bulk2}
\end{equation}
Equations (\ref{bulk1}), (\ref{bulk2}) and (\ref{3.3}) determine
the brane and bulk geometry completely. They are the main result
of this paper. Below we will discuss mainly its application to the
scenario of Ref.\cite{carroll-foot}. Before that, two remarks are
in order about those solutions.

First, a whole new feature of the solution (\ref{bulk2}) compared
to the Einstein case is that the deficit angle in the extra
dimensions will now depend on the geometry of the branes. From
this, we can find an interesting geometric argument in favor of a
positive Gauss-Bonnet coupling constant. In the case of a negative
Gauss-Bonnet coupling constant, the geometry will become singular
when $\Lambda_4>-1/(4\alpha)$. Since we expect $\Lambda_4$ to be
very large during the inflation era, the requirement of a
nonsingular geometry forces us to rule out a negative Gauss-Bonnet
coupling constant.

Second, the brane geometry in 6-dimensional Einstein-Gauss-Bonnet
gravity is also discussed in Ref.\cite{bgns}. The authors actually
considered only the longitudinal component of the EGB equation and
concluded that Einstein gravity will restore on the brane. Due to
our analysis, the brane geometry is determined by the transverse
component of the EGB equation and while the longitudinal equation
looks like Einstein equation, it actually determines the bulk
geometry \emph{after} the brane geometry is found by the
transverse equation. This can be seen more clearly by the
expression for the 4-dimensional cosmological constant in
Ref.\cite{bgns} (Eq.(21) in that reference). Actually, Eq.(21) in
Ref.\cite{bgns} is exactly Eq.(\ref{bulk2}), from which we can see
that it actually determines the deficit angle \emph{after} the
4-dimensional cosmological constant is found from Eq.(\ref{3.3}).

Now, let's discuss a specific example of the solutions
(\ref{bulk1}), (\ref{bulk2}) and (\ref{3.3}). A lot of the recent
works on codimension 2 branes are motivated by the exact solution
presented by Carroll and Guica \cite{carroll-foot} which shows
explicitly the independence of the 4-dimensional geometry on the
brane tension. Thus we think it is most important to discuss the
corresponding solutions in EGB gravity and compare it with that of
Ref.\cite{carroll-foot}. The solution presented by Carroll and
Guica assumes a bulk cosmological constant and a magnetic flux,
which is described by the bulk action \cite{carroll-foot},
\begin{equation}
  S_6 = \int d^6X\sqrt{|G|}\, \left( {1\over 2}M_6^4 R - \lambda
  - {1\over 4} F_{AB} F^{AB}\right)\ ,
  \label{action}
\end{equation}
where $M_6$ is the 6-dimensional reduced Planck mass and $\lambda
$ is the 6-dimensional vacuum energy density. The 2-form field
strength takes the form $F_{ab}=\sqrt{|\gamma|}B_0\epsilon_{ab}$,
where $B_0$ is a constant and $\epsilon_{ab}$ is the standard
antisymmetric tensor. Other components of $F_{AB}$ vanish
identically. This model is originally suggested to stabilize the
extra dimensions \cite{sundrum, sta}.

The bulk energy-momentum tensor contains contributions from both the
bulk cosmological constant and the gauge field,

\begin{equation}
  T^B_{AB} = T_{AB}^{\lambda} + T_{AB}^{F}\ ,\label{1.4}
\end{equation}
for which the explicit forms are
\begin{eqnarray}
  T_{AB}^{\lambda} &=& -\lambda\left(\begin{matrix} g_{\mu\nu} & 0 \cr
  0 & \gamma_{ab} \end{matrix}\right)\cr
  T_{AB}^{F} &=& -{1\over 2} B_0^2\left(\begin{matrix} g_{\mu\nu} & 0 \cr
  0 & -\gamma_{ab} \end{matrix} \right)\label{1.3} .
\end{eqnarray}
So we have $T^B_1=-4\lambda-2B_0^2$ and $T^B_2=-2\lambda+B_0^2$.

At first, we generalize the flat brane solution of
Ref.\cite{carroll-foot} to de Sitter brane, which is given by
\begin{equation}
a_0^2=\frac{M_6^4}{2\lambda-3M_6^4\Lambda_4}\ ,\label{3.71}
\end{equation}
\begin{equation}
 \beta = 1-{\sigma\over
2\pi M_6^4}\ ,\label{3.7}
\end{equation}
\begin{equation}
M_6^4\Lambda_4=\frac{1}{2}\lambda-\frac{1}{4}B_0^2\ .\label{3.8}
\end{equation}

It is interesting to note that for the geometry to be nonsingular,
from Eq.(\ref{3.71}), we must have $\Lambda_4<2\lambda/(3M_6^4)$.
However, from Eq.(\ref{3.8}), this is always satisfied. Thus the
de Sitter geometry of the brane will never make the bulk geometry
singular.

From Eq.(\ref{3.8}), we can see that the puzzle of a small
4-dimensional cosmological constant is now transformed to the
question of explaining a fine-tuning between the 6-dimensional
vacuum energy and the magnetic flux, which is a purely bulk
problem. Thus in this scenario the cosmological constant problem
is moved completely into the bulk. Of course, this does not solve
the cosmological constant problem, but it transforms the nature of
the problem in an interesting way. At a first glance, it is
tempting to appeal to the usual supersymmetry argument
\cite{carroll-cc} to set both $\lambda$ and $B_0^2$ very small,
thus avoiding fine-tuning between them. However, this cannot work.
From Eq.(\ref{3.71}), we can see that we must require either
$\lambda$ or $B_0^2$ to be of the order $M_6^6$ so that the size
of the extra dimensions can be phenomenologically viable. Thus,
there is a real fine-tuning problem in the bulk. Currently, we
still do not know whether this fine-tuning can be technically
natural. Thus, it would be very interesting that if in the EGB
gravity, we can have a way to release this fine-tuning. We will
see below that when the Gauss-Bonnet coupling constant is large,
this is possible.

Then, we turn to the discussion of solutions in EGB gravity. From
Eqs.(\ref{bulk1}), (\ref{bulk2}) and (\ref{3.3}), the
corresponding solution in EGB gravity is given by
\begin{equation}
a_0^2=\frac{M_6^4(1+4\alpha\Lambda_4)}{2\lambda-3M_6^4\Lambda_4-{4\over3}\alpha
M_6^4\Lambda_4^2}\ , \label{result1}
\end{equation}
\begin{equation}
\beta=1-\frac{\sigma}{2\pi M_6^4(1+4\alpha\Lambda_4)}\
,\label{result2}
\end{equation}
\begin{equation}
\Lambda_4={3\over4\alpha}\left(\sqrt{1+{2\alpha\over3}{2\lambda-B_0^2\over
M_6^4}}-1\right)\ .\label{4dcc}
\end{equation}

As a first remark, while we have discussed above, a negative
$\alpha$ may result in a singular spacetime; for a positive
$\alpha$, while it is not very obvious, it still can be shown that
the geometry is always nonsingular by an argument that is similar
to the Einstein case.

Then let's discuss the cosmological constant problem in EGB
gravity as expressed by Eq.(\ref{4dcc}). Since we generally have
$\lambda<M_6^6$ and $B_0^2<M_6^6$, so when $\alpha M_6^2< 1$, we
have $\alpha(2\lambda-B_0^2)/M_6^4\ll 1$. Expanding the RHS of
Eq.(\ref{4dcc}) to first order, we can find that $\Lambda_4\sim
(2\lambda-B_0^2)/M_6^4$. Thus, in this case we are actually facing
the same fine-tuning as in Einstein gravity in order to get a
small cosmological constant. This is not a surprise, since it is
natural for the solution to reduce to the Einstein case when
$\alpha$ is small. So what is interesting is the case where the
Gauss-Bonnet coupling constant is large. Let's assume
$2\lambda-B_0^2\sim M_6^6$, i.e. we do not have a fine-tuning in
the bulk, and when $\alpha M_6^2\gg 1$, i.e. considering the case
of a large Gauss-Bonnet coupling constant, from Eq.(\ref{4dcc}),
we can obtain
\begin{equation}
\Lambda_4\sim{M_6\over\sqrt{\alpha}}\ .\label{3.10}
\end{equation}
Thus even if we do not have a fine-tuning in the bulk, for a
sufficiently large $\alpha$, we still can get a small
4-dimensional cosmological constant. In this case, the current
cosmological expansion acceleration is actually driven by the
6-dimensional Gauss-Bonnet term, which is in some sense similar to
the recent model of $1/R$ gravity proposed by Carroll \emph{et
al.} \cite{carroll-1/R}: the current cosmological expansion
acceleration is driven by a $1/R$ term in the 4-dimensional
gravitational lagrangian. Of course, in order for the $\Lambda_4$
to be the order of the observational value, $\alpha^{-1}$ also
needs to be fine-tuned to an extremely small value. Thus in the
current case, we have actually traded the fine-tuning in the bulk
to a fine-tuning in the Gauss-Bonnet coupling constant. Although
this new fine-tuning also seems unnatural now, the cosmological
constant problem is so hard to solve that it is worth transforming
it to a new problem for further investigations. Furthermore, this
shows the qualitative feature of what will happen if we consider
higher-derivative gravity in the bulk. Maybe considering more
complicated higher-derivative gravity theories, such as forth
order combinations of the curvature tensor can further release the
fine-tuning in a more natural way. This deserves further
investigating. It is worth mentioning that similar fine-tuning
problem also happens in the $1/R$ gravity: the coefficient of the
$1/R$ term should also be extremely small to account for the
current cosmic accelerating expansion \cite{carroll-1/R}. In the
$1/R$ gravity, this is unnatural from an effective field point of
view and can lead to some inconsistencies when the theory is
treated quantum mechanically \cite{flanagan}. Now we still do not
know whether similar problem will be presenting here.

As a final remark, in Ref.\cite{navarro}, Navarro considered using
a 4-form field in place of the 2-form field in the action
(\ref{action}). By using Eqs.(\ref{bulk1}), (\ref{bulk2}) and
(\ref{3.3}), it is trivial to generalize Navarro's solution to the
EGB gravity.

\section{Conclusions and Discussions}
\label{CD}

In this paper, we have discussed the gravitational properties of
codimension 2 branes in Einstein-Gauss-Bonnet gravity and their
implications in addressing the cosmological constant problems.

Although the current scenario is originally introduced to discuss
the cosmological constant problem, it is also mandatory that
cosmological models from String Theory should be reconciled with
inflation,  now a quite well-established ingredient of modern
cosmology \cite{carroll-cos} (see, e.g., Ref.\cite{lidsey} for a
recent review of braneworld inflation; inflation in 5-dimensional
EGB gravity is recently discussed in Ref.\cite{gbinflation}). When
considering inflation model building in the present scenario, an
observation is that the inflaton must be a bulk field. This is in
sharp contrast to the discussions of the codimension 1 case, where
most of the inflation model assumes the inflaton to be confined on
the brane \cite{lidsey}. The reason for this is simple. Current
observation of the CMB power spectrum tells us that during
inflation, the energy density of inflaton should be almost
constant \cite{carroll-cos}. Thus, if the inflaton is a field
confined on the brane, then during inflation it will behave just
like the brane tension. So the above analysis tells us that it
cannot affect the 4-dimensional geometry. On the other hand, if
the inflaton is a bulk field, then its effects during inflation
are just equivalent to a renormalization of the 6-dimensional
cosmological constant $\lambda$.

So in the Einstein gravity case, the Hubble parameter during
inflation $H^2\equiv\Lambda_4/3$ will be given from Eq.(\ref{3.8})
by
\begin{equation}
H^2\sim V/M_6^4\ ,\label{4.1}
\end{equation}
where $V$ is the potential of the bulk inflaton field. Thus the
energy scale of the potential would be of order $(H/M_6)^{1/3}M_6$
during inflation. In the original model of large extra dimensions
\cite{add}, in order to address the gauge hierarchy problem, the
6-dimensional reduced Planck mass is assumed at most a few orders
higher than the supersymmetry breaking scale which is of order 1
TeV. On the other hand, current CMB data prefers a high inflation
scale which is at most several orders of magnitude smaller than
the GUT scale $\sim 10^{16}$ GeV \cite{carroll-cos}. Thus, the
potential $V$ during inflation is necessarily larger than $M_6$.
If the inflaton is a brane field, there is nothing unnatural here.
But as we have commented above, inflaton must be a bulk field now.
So it is very unnatural for a bulk field to have an energy scale
larger than the bulk Planck mass. Therefore, implementing
successful inflation scenario encounters fundamental difficulties
in codimension 2 brane scenarios \footnote{If considering warped
extra dimensions, it is possible to get any value of 4-dimensional
cosmological constant by tuning the warp factor without a large
potential in the bulk \cite{cline}, thus avoiding the problem
discussed above. We thank J. Vinet for pointing this possibility
to us. However, in this case, 4-dimensional cosmological constant
is actually only an integration constant and cannot be determined
from the model parameter. We think this seems not a very good
feature so we do not consider this possibility further in this
work while this possibility of course deserves further detailed
considerations.}.

The situation is worse in EGB gravity. From Eq.(\ref{4dcc}), the
Hubble parameter during inflation will be given by Eq.(\ref{4.1})
when $M_6^2\alpha\ll 1$ and it reduces to the Einstein case. When
$M_6^2\alpha\gg 1$, from Eq.(\ref{3.10}), the Hubble parameter
will be given by
\begin{equation}
H^2\sim {M_6\over\sqrt{\alpha}}\sqrt{V}\label{}
\end{equation}
Thus the energy scale of the potential would be of order
$(H/M_6)^{2/3}(\alpha M_6^2)^{1/6}M_6$ during inflation. Then a
higher potential energy is needed compared with the Einstein case
(\ref{4.1}) to implement the inflation. This makes the problem we
discussed above more severe.

Faced with the above problem, it is worth considering other
mechanism of driving an inflation on the brane rather than a bulk
scalar field. A seemingly promising candidate is the $R^2$
inflationary model of Starobinsky \cite{star}. However, in order
to avoid the above problem, we assume that the $R^2$ term is only
induced on the brane, like the case of induced gravity model given
by Dvali \emph{et al.} \cite{dvali}. More concretely, we may
consider adding to the bulk lagrangian (\ref{action}) an induced
$R^2$ term,
\begin{equation}
S_{induced}=\int d^4x\sqrt{|g|}\tilde{\alpha} R[g]^2\ ,\label{}
\end{equation}
where $\tilde{\alpha}$ will be of order $M_4^{-2}$ \cite{star}. It
is worth commenting that such a term may be induced by quantum
effects of conformal fields on the brane, and $R^2$ inflation on
codimension 1 braneworld has been discussed in
Ref.\cite{odintsov-r2}. This and other possibilities to handle the
inflation model building problems deserve further investigation.

\section*{Acknowledgments}
PW would like to thank Sergei D. Odintsov,  Jeremie Vinet and Liu
Zhao for helpful comment on the manuscript. XHM would also like to
express his thanks to the Physics Department of UoA for its
hospitality extended to him. This work is supported partly by an
ICSC-World Laboratory Scholarship, a China NSF and Doctoral
Foundation of National Education Ministry.

\end{document}